\documentclass{ws-rv975x65}
\usepackage{subfigure}     
\usepackage{ws-rv-thm}     
\usepackage[square]{ws-rv-van}     
\usepackage{graphicx}
\usepackage{wasysym}
\usepackage[hyperindex,breaklinks]{hyperref}

\makeindex
 \RequirePackage{lineno}

\begin{document}
\chapter[Multi-messenger astronomy]{The dawn of multi-messenger astronomy}\label{multimessenger}

\author[M. Santander]{Marcos Santander}
\address{Barnard College, Columbia University \\
3009 Broadway, New York, NY, USA \\
santander@nevis.columbia.edu}

\begin{abstract}
The recent discoveries of high-energy astrophysical neutrinos and gravitational waves have opened new windows of exploration to the Universe. Combining neutrino observations with measurements of electromagnetic radiation and cosmic rays promises to unveil the sources responsible for the neutrino emission and to help solve long-standing problems in astrophysics such as the origin of cosmic rays. Neutrino observations may also help localize gravitational-wave sources, and enable the study of their astrophysical progenitors. In this work we review the current status and future plans for multi-messenger searches of neutrino sources. 
\end{abstract}
\body

\section{Introduction}\label{sec:intro}

The continuing study of the high-energy sky has revealed a large number of powerful astrophysical objects capable of emitting radiation across the entire electromagnetic spectrum. 
The high magnetic fields and strong astrophysical shocks observed in some of these objects are expected to be responsible for the acceleration of cosmic rays up to the highest observable energies, in the $10^{20}$ eV range. As they are being accelerated, or during their propagation, cosmic rays can interact with ambient material or radiation fields, leading to the production of high-energy neutrinos and gamma rays through the decay of charged and neutral mesons~\cite{Kelner:2006tc, Kelner:2008ke}. Fast transient sources which are potential cosmic-ray accelerators, such as gamma-ray bursts (GRBs), may also emit gravitational waves due to bulk motion of matter in the source progenitor.

A complete understanding of the most energetic phenomena in the Universe therefore calls for a joint study of the different ``cosmic messengers'' they emit: cosmic rays, neutrinos, photons, and gravitational waves. Multi-messenger astronomy is an emerging subfield of high-energy astrophysics that aims at combining observations from instruments sensitive to these different messenger particles. This approach may solve the long-standing mystery of the sources of cosmic rays, further our understanding of particle acceleration in astrophysical shocks, increase the sensitivity of searches for gravitational-wave emitters, and potentially unveil new types of sources through the detection of spatial and temporal correlations between two or more messenger channels.

A major result for the field has been the detection of an astrophysical flux of neutrinos by the IceCube observatory at the level of $ E^2_{\nu} J_{\nu}(E_{\nu}) \sim 10^{-8}$ GeV cm$^{-2}$ s$^{-1}$ sr$^{-1}$ per neutrino flavor in the energy range between a few tens of TeV and a few PeV~\cite{HESE3yr}. While no point sources have been detected so far, the flux upper limits set by IceCube and ANTARES searches~\cite{IC4yearPS} are at the level of 1--10\% of the all-sky astrophysical flux, hinting at a large population of neutrino sources. No significant correlation has been found with the Galactic Plane, which tends to favor an extragalactic origin of the astrophysical neutrinos, with a potential sub-dominant contribution from galactic sources.

\begin{figure}[htb]
\center
\raisebox{-0.5\height}{\includegraphics[width=1\textwidth]{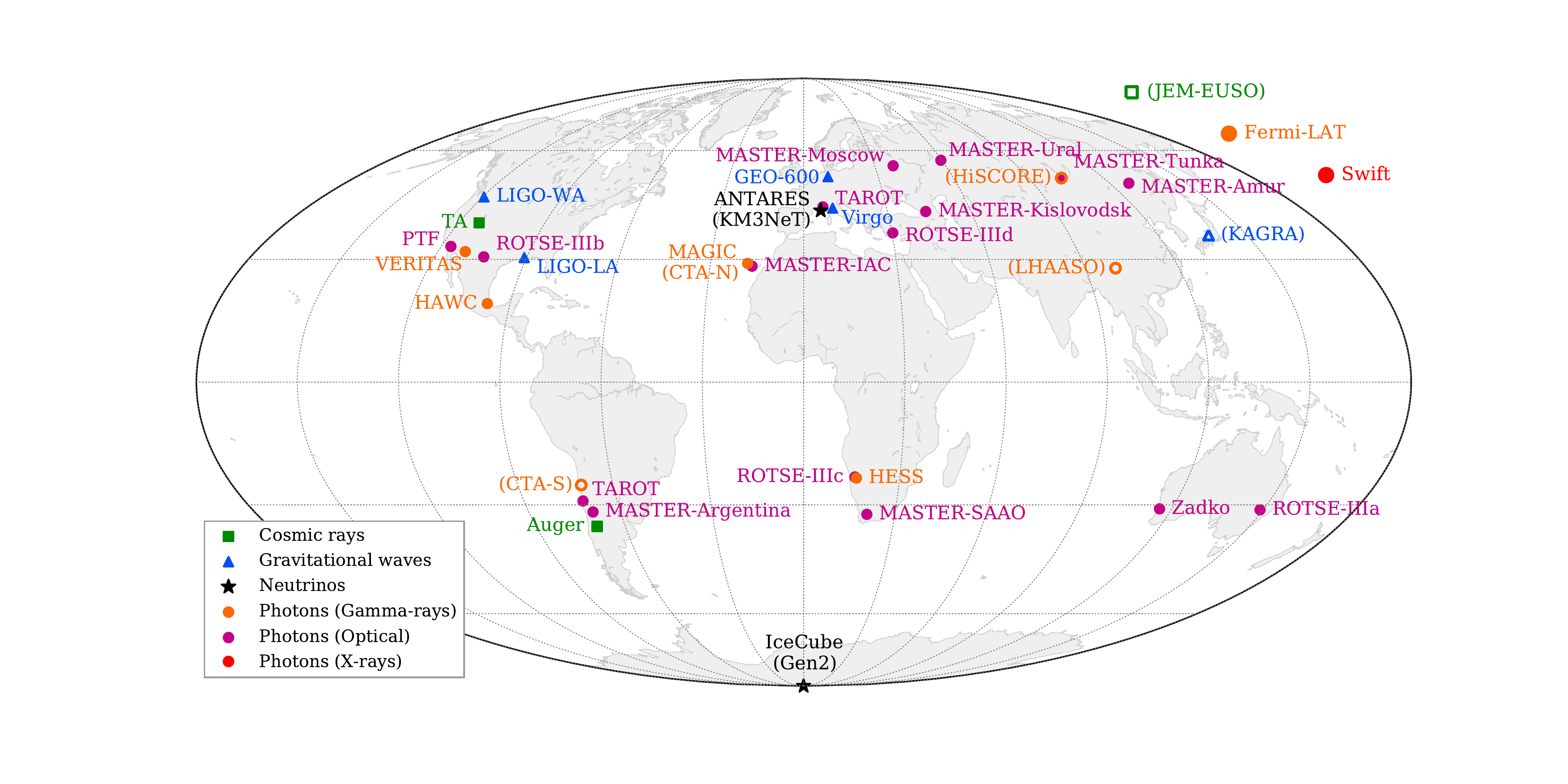}}
\caption{Global distribution of some of the observatories (past, present, and future) dedicated to multi-messenger studies. Current and past facilities are shown with filled markers. Some representative future detectors are shown with empty markers with their names in parentheses. }
\label{fig:mm_map}
\end{figure}

The search for neutrino sources should benefit from the boost in sensitivity provided by the multi-messenger approach. Over the last few years, data from a large network of astrophysical observatories around the world (shown in Fig.~\ref{fig:mm_map}) has been used to study correlations between neutrino events and other messenger signals. As more facilities go online in coming years, a drastic increase in the sensitivity of these searches is expected, which may finally reveal these elusive sources. 

In the following sections, we review recent results from multi-messenger searches for high-energy neutrino sources. Searches for electromagnetic counterparts to the neutrino emission are described in Section~\ref{sec:photons}. Section~\ref{sec:cosmic-ray} covers results from correlation studies between neutrinos and high-energy cosmic rays. In Section~\ref{sec:gw} we describe joint searches of neutrino and gravitational-wave emitters. We describe current efforts to build a multi-messenger observatory network to provide rapid communications of high-energy astrophysical events in Section~\ref{sec:transient}. We present conclusions and an outlook for future studies in Section~\ref{sec:conclusions}.
\section{Photons}\label{sec:photons}

The high-energy hadronic interactions responsible for the creation of the astrophysical neutrinos observed by IceCube should also produce gamma rays through the decay of neutral pions. Lower energy photons may also be detectable for those cases where the source or the propagation medium is opaque to gamma rays. Studies of spatial and temporal correlations between neutrinos and EM radiation rely on searching for neutrino emission from known EM sources where cosmic-ray acceleration is expected, or on performing follow-up EM observations of high-energy neutrino positions that are likely astrophysical in origin. The first approach has been used to set upper limits on the neutrino flux from GRBs, AGNs, SNRs, and galaxy clusters. In the case of GRBs, null neutrino detections have set important constraints on models that postulate these objects as sources of cosmic rays with energies above $10^{18}$ eV (see \cite{2015ApJ...805L...5A} for a recent update on this work). In the following subsections, we summarize recent EM--$\nu$ searches that use the second approach. 

While temporal correlations can be explored down to the microsecond level (given the precision in the determination of the neutrino arrival time), the main challenge for spatial correlation studies is presented by the limited angular resolution of neutrino directional reconstructions. In IceCube, the angular resolution of ``cascade'' events produced by charged-current $\nu_{e, \tau}$, or neutral-current interactions of any flavor, is about $15^{\circ}$ at energies above 100 TeV. Charged-current $\nu_{\mu}$ interactions produce km-long muon ``tracks" that can be typically reconstructed to within $1^{\circ}$. The muon reconstruction capability of IceCube has been validated by the observation of the cosmic-ray shadow of the Moon within $0.2^{\circ}$ of its expected position~\cite{I3MoonShadow}. Deep-sea neutrino telescopes benefit from the longer scattering length of Cherenkov light in water to reconstruct neutrino events to higher precision. The angular resolution of ANTARES~\cite{AdrianMartinez:2012rp} for muon tracks is believed to be better than $0.3^{\circ}$ above 10 TeV and about $4^{\circ}$ for cascade events, while for the future KM3NeT detector~\cite{Margiotta:2014gza} the angular resolution for cascades is expected to improve to $2^{\circ}$~\cite{KM3NetCascades}. Recent correlation studies have concentrated on searching for EM emission coincident with muon track positions to benefit from the better angular resolution of these events and reduce the probability of accidental correlations. 

\subsection{Gamma-rays}

At production, the flux of TeV--PeV astrophysical neutrinos should be associated with a flux of gamma-rays of similar spectral characteristics. Photons in this energy range are attenuated during propagation by pair-producing on background radiation fields, with the extragalactic background light (EBL) dominating at $E_{\gamma} <$ 10 TeV and the cosmic microwave background (CMB) for $E_{\gamma} > $ 10 TeV. At PeV energies, the photon attenuation length is below 10 kpc, which restricts correlation studies of PeV photons and neutrinos to our galaxy. Past gamma-ray observations have been used to test the association of the astrophysical neutrinos with the Galactic Plane~\cite{Ahlers:2013xia, Kistler:2015oae}, the Galactic Halo \cite{Taylor:2014hya}, and the \emph{Fermi} ``bubbles''~\cite{Lunardini:2015laa}. The sensitivity of these tests will be greatly improved by observations from current and future air-shower arrays, such as IceTop~\cite{IceCube:2012nn}, HAWC~\cite{Abeysekara:2013tza}, LHAASO~\cite{Cui:2014bda} and HiScore~\cite{Tluczykont:2014cva}. 

\begin{figure}[htb]
\centerline{\includegraphics[width=0.8\textwidth]{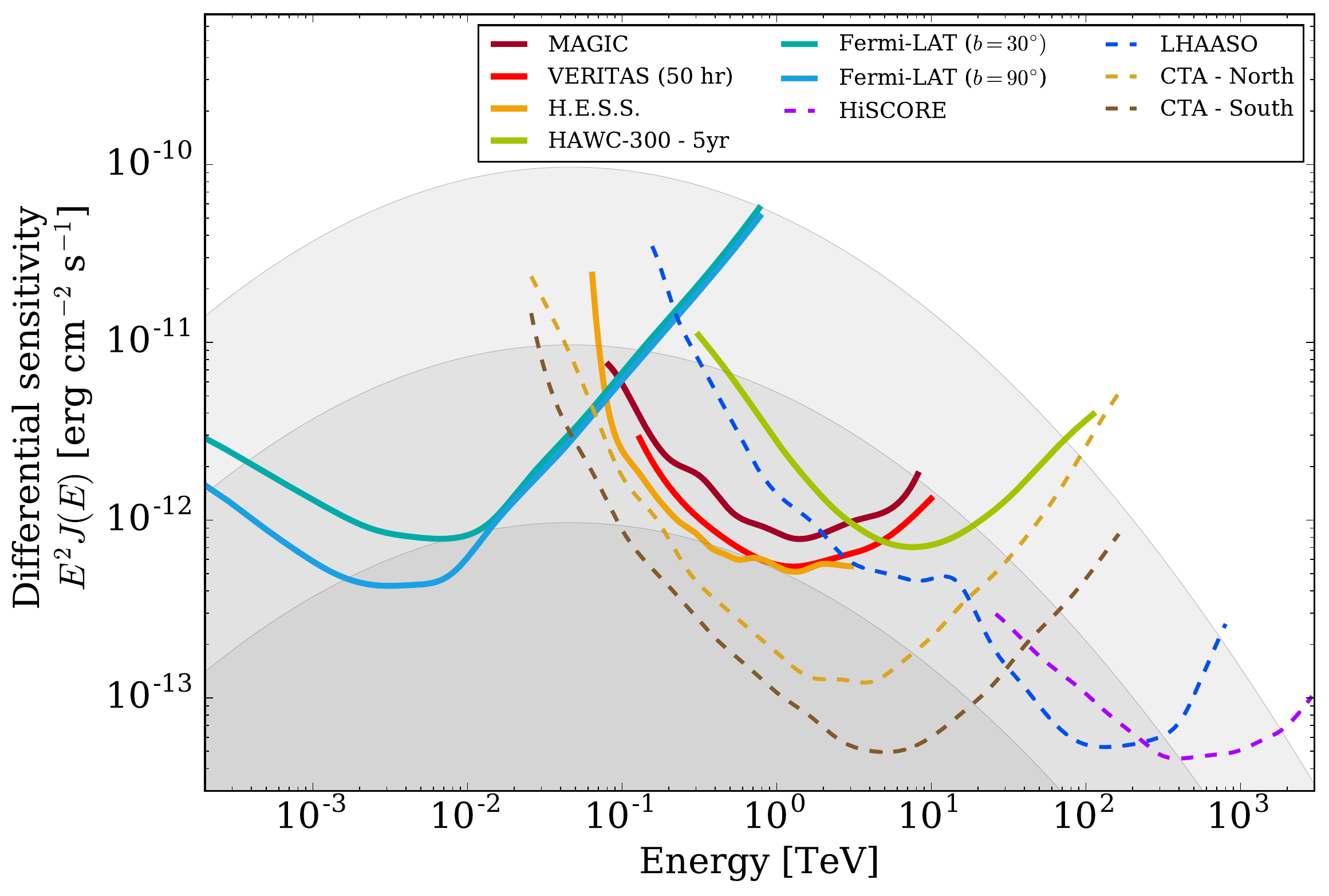}}
\caption{Differential $5\sigma$ sensitivity of current (solid lines) and future (dashed lines) gamma-ray observatories. The \emph{Fermi}-LAT~\cite{FermiLATPerformance, Atwood:2013rka} sensitivity curve is given for a 10 year exposure at two galactic latitudes ($30^{\circ}$ and $90^{\circ}$). The \emph{Fermi}-LAT and HAWC curves are given for quarter-decade energy bins. The VERITAS~\cite{Park:2015ysa}, MAGIC~\cite{2016APh....72...76A}, H.E.S.S.~\cite{Holler:2015tca}, and CTA~\cite{CTASensitivity} curves are given for 50 hours of observation and 5 energy bins per decade. The HAWC 300 sensitivity~\cite{Abeysekara:2013tza}, and that of the future HiScore~\cite{Tluczykont:2014cva} and LHAASO~\cite{LHAASO:ICRC} arrays, is given for a five-year exposure. For reference, the shaded grey regions indicate, from the top, 100\%, 10\%, and 1\% levels of the gamma-ray spectrum of the Crab nebula.
} 
\label{fig:sensitivities}
\end{figure}

Neutrino correlations with sources of extragalactic gamma-rays can be investigated at GeV--TeV energies, where absorption is not as severe, if the hadronic gamma emission extends to this energy range. The main instruments in this band are the \emph{Fermi} Large Area Telescope (LAT)~\cite{2009ApJ...697.1071A}, the H.E.S.S.~\cite{Aharonian:2006pe}, MAGIC~\cite{2013arXiv1309.6979I}, and VERITAS ~\cite{2002APh....17..221W} ground-based telescopes, and the HAWC array. The sensitivities of current and future gamma-ray telescopes are shown in Fig.~\ref{fig:sensitivities}. 

The connection between the neutrino flux and extragalactic radiation backgrounds has been explored in recent studies. Simple extrapolations of the astrophysical neutrino flux down to GeV energies lead to an associated photon flux that can account for a significant fraction or even overflow (depending on the assumed neutrino spectral index) the isotropic gamma-ray background (IGRB) measured by \emph{Fermi}-LAT~\cite{Ackermann:2014usa}. However, \emph{Fermi} source population studies~\cite{DiMauro:2015tfa} indicate that the IGRB is dominated by unresolved AGNs (typically assumed to be leptonic sources) which results in a lower fraction of the IGRB that could be connected to neutrinos. While significant uncertainties remain on these extrapolations,  current measurements are starting to probe the role that proton-proton sources (such as the archetypical star-forming galaxies) play in diffuse gamma-ray backgrounds~\cite{Ando:2015bva, Bechtol:2015uqb}. 

\begin{figure}[tb]
\begin{center}
\raisebox{-0.5\height}{\includegraphics[width=0.35\textwidth, angle=90,origin=c]{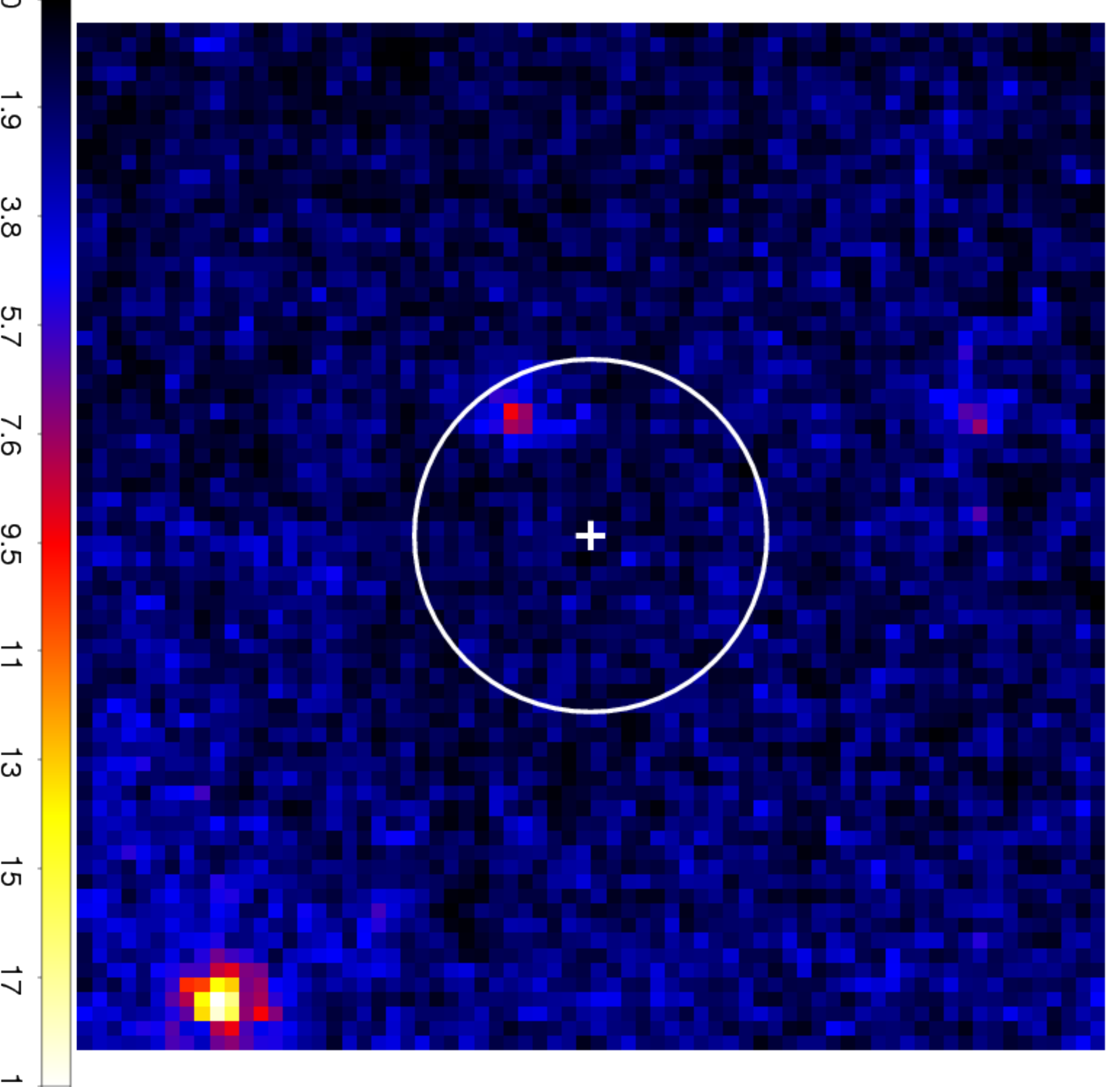}}
\raisebox{-0.5\height}{\includegraphics[width=0.56\textwidth]{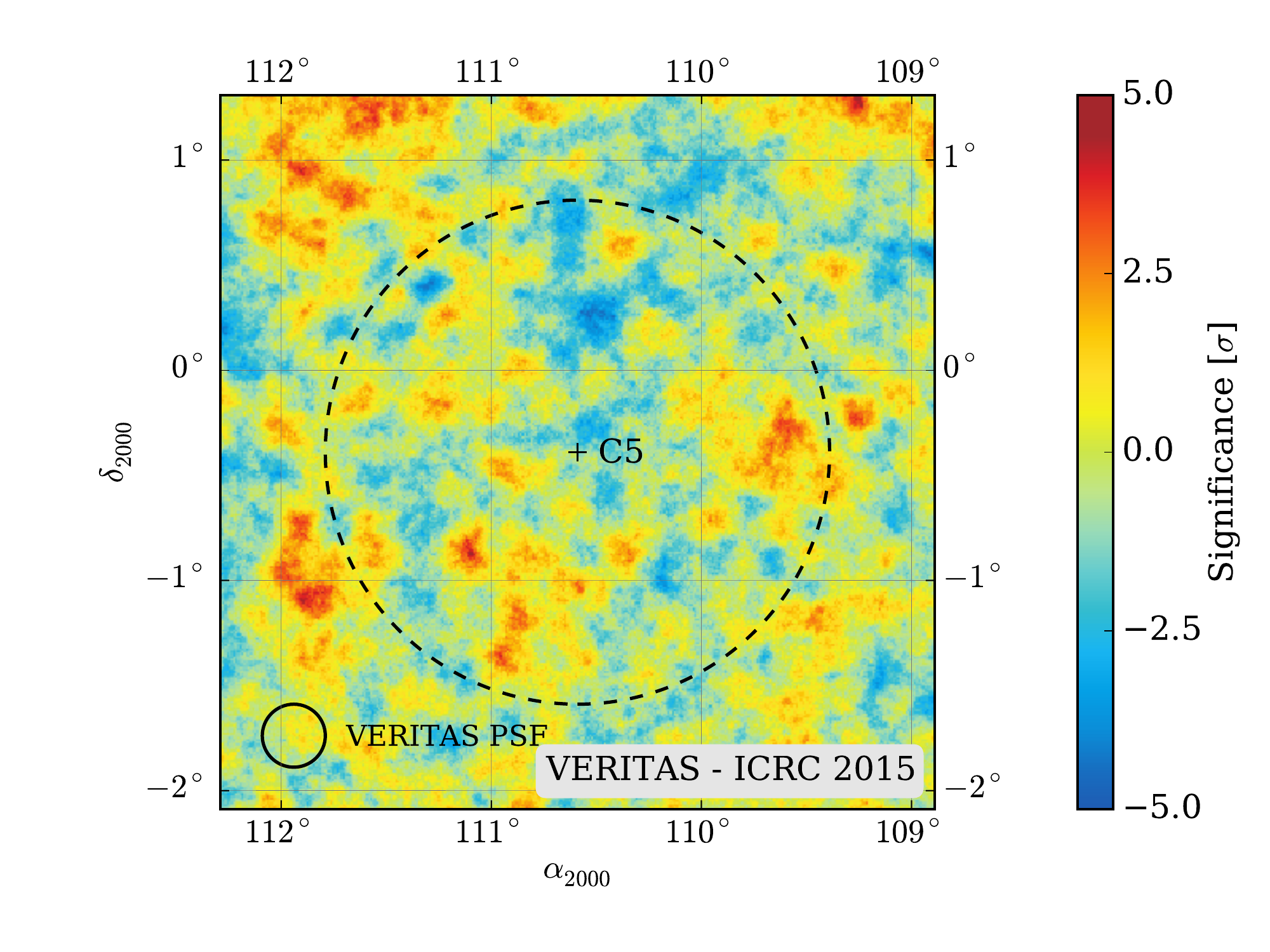}}
\end{center}
\caption{ Gamma-ray observations of an IceCube muon-track position (event \#5 in \cite{HESE3yr}). \emph{Left}: \emph{Fermi}-LAT test-statistic sky map of gamma rays in the 1-300 GeV range (see \cite{Brown21072015}). \emph{Right:} VERITAS significance sky map (see \cite{2015arXiv150900517S}). 
The white circle in the \emph{Fermi} sky map and the dashed circles in the VERITAS map show the $<1.2^{\circ}$ median angular resolution claimed for this neutrino event.}
\label{fig:icecube_id5}
\end{figure}

The large field of view and high duty cycle of the LAT provides temporal and spatial gamma-ray coverage for a large fraction of the neutrino events detected by IceCube. Data from the LAT has been analyzed to search for new sources, or flux enhancements in known ones, in coincidence with IceCube neutrino events. No spatially-coincident gamma emission has been found in correlation with muon track events~\cite{Brown21072015} with the exception of a neutrino candidate event near the location of the gamma-ray blazar PKS 0723-008 (see Fig.~\ref{fig:icecube_id5}), likely due to a chance alignment ($p$ = 37\%). It has been recently reported~\cite{LATBigBird} that a 2 PeV cascade event detected by IceCube occurred in relative temporal and spatial coincidence with an extended high-fluence flare of the blazar PKS B1424-418, although also in this case the association does not appear to be statistically significant ($p$ = 5\%).

Searches for neutrino gamma-ray counterparts in the very-high-energy range (VHE, $E_{\gamma} >$  100 GeV) are underway using the H.E.S.S., MAGIC, and VERITAS Imaging Air Cherenkov Telescopes (IACTs) and the HAWC air shower array. For IACTs, these searches are limited to the observation of muon track positions, given the 3.5$^{\circ}$--5$^{\circ}$ field-of-view of current generation telescopes. The VERITAS and H.E.S.S. telescopes have observed muon track positions published by ANTARES~\cite{Schussler:2013lja} and IceCube~\cite{2015arXiv150900517S, 2015arXiv150903035S}, and null results from these studies have set constraints on the steady VHE gamma-ray flux associated with each neutrino position. Increasing the sensitivity to transient gamma-ray sources requires a system that can issue prompt alerts to VHE instruments if a hint of an increase in neutrino activity is detected. Since 2012, IceCube operates a program that sends triggers to MAGIC and VERITAS whenever the number of neutrino events detected over a certain period around a VHE source crosses a predefined significance threshold~\cite{2013arXiv1309.6979I}, so far with no significant gamma-ray detections.

A golden channel for follow-up observations is the sample of high-energy through-going muon events used by IceCube to measure the astrophysical $\nu_{\mu}$ flux~\cite{Aartsen:2015rwa} given its good angular resolution and high astrophysical purity. At energies above a few hundred TeV, where the astrophysical neutrino flux dominates the atmospheric background, the IceCube effective area to through-going muon neutrinos is more than ten times larger than for ``starting'' neutrino events, where the first interaction occurs in the detector volume. The highest energy neutrino detected so far (recently reported by IceCube~\cite{PeVmuonATel}) comes from this sample and has an energy of $2.6 \pm 0.3$ PeV with an atmospheric \emph{p}-value of $< 0.01\%$. An archival analysis of HAWC data around the time of the event showed no gamma-ray emission at the neutrino location. Current efforts are underway to promptly circulate the positions of high-energy starting and through-going muon neutrino events to partner instruments using the AMON network (see Section~\ref{sec:transient}), which would boost the sensitivity of EM follow-ups to transient neutrino sources. Future searches for VHE gamma-ray neutrino counterparts will receive a significant boost from the construction of the Cherenkov Telescope Array (CTA)~\cite{Acharya:2013sxa}, which will provide an order-of-magnitude improvement in sensitivity with respect to current IACTs (see Fig.~\ref{fig:sensitivities}). At MeV energies, proposed missions such as the ComPair satellite~\cite{Moiseev:2015lva} can provide a large field-of-view coverage of the sky with an angular resolution in the 1$^{\circ}$ to 10$^{\circ}$ range.

\subsection{Multi-wavelength observations}

Besides gamma rays, other wavelength bands are being explored to search for transient EM emission associated with neutrino events. Realtime alerts from ANTARES and IceCube are currently sent to a network of optical and X-ray telescopes to search for transient sources such as GRBs and core-collapse supernovae (CCSNe) in correlation with interesting neutrino positions. 
Since 2008, IceCube operates optical (OFU, \cite{Abbasi:2011ja}) and X-ray (XFU) follow-up programs in parallel to the gamma-ray follow-up program described in the previous subsection. The rate of false trigger alerts is reduced by requiring that two or more spatially-coincident neutrinos are detected within a certain time window~\cite{Kowalski:2007xb}. OFU alerts have been sent to the ROTSE telescope network (which has since stopped operations) and the Palomar Transient Factory (PTF), and have been supplemented by retrospectives searches through the Pan-STARRS1 $3\pi$ survey data. The XFU program triggers observations in the 0.3-10 keV band using the XRT X-ray instrument onboard the \emph{Swift} satellite.  

\begin{figure}[htb]
\center
\raisebox{-0.5\height}{\includegraphics[width=0.34\textwidth]{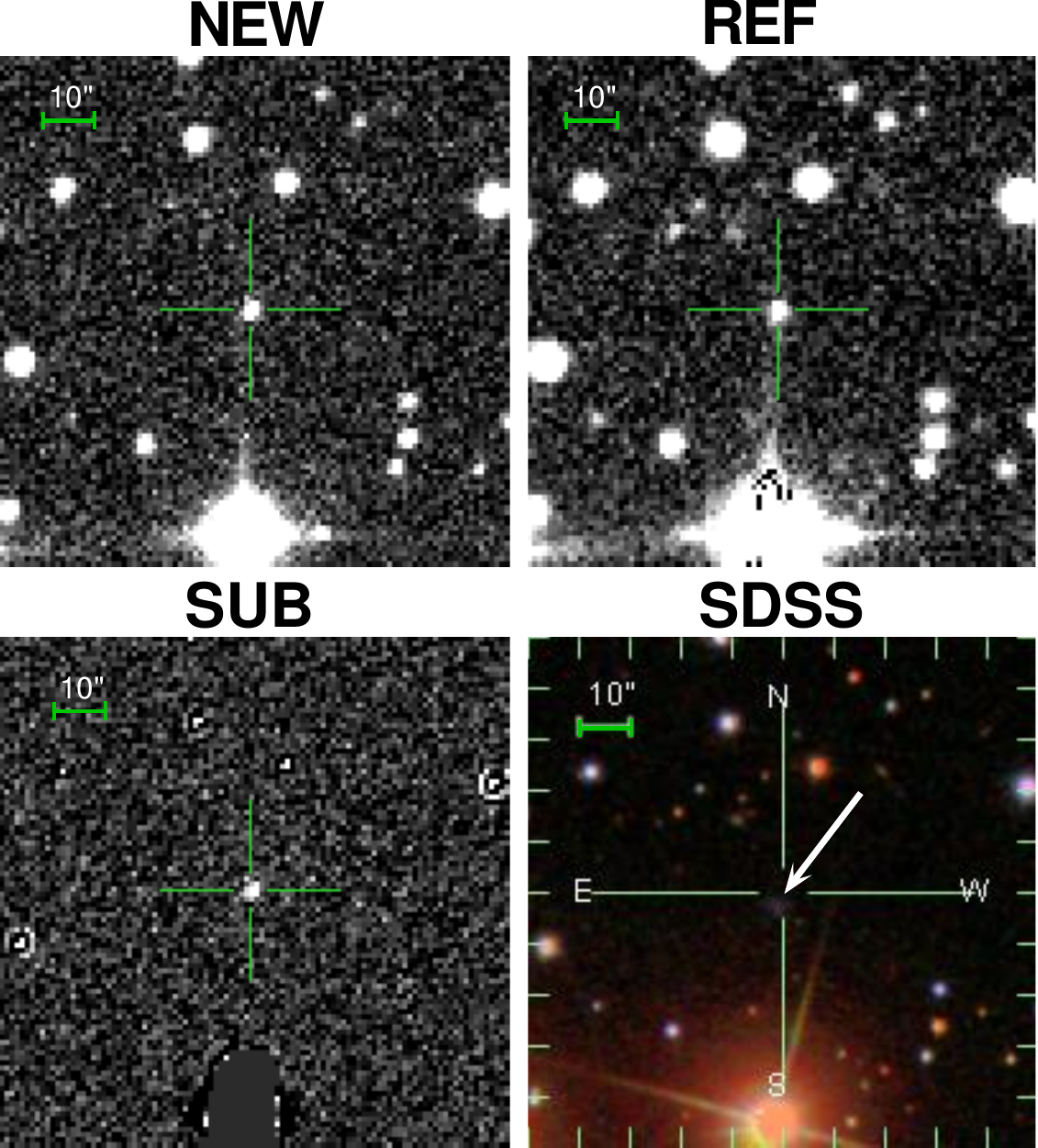}}
\raisebox{-0.6\height}{\includegraphics[width=0.45\textwidth]{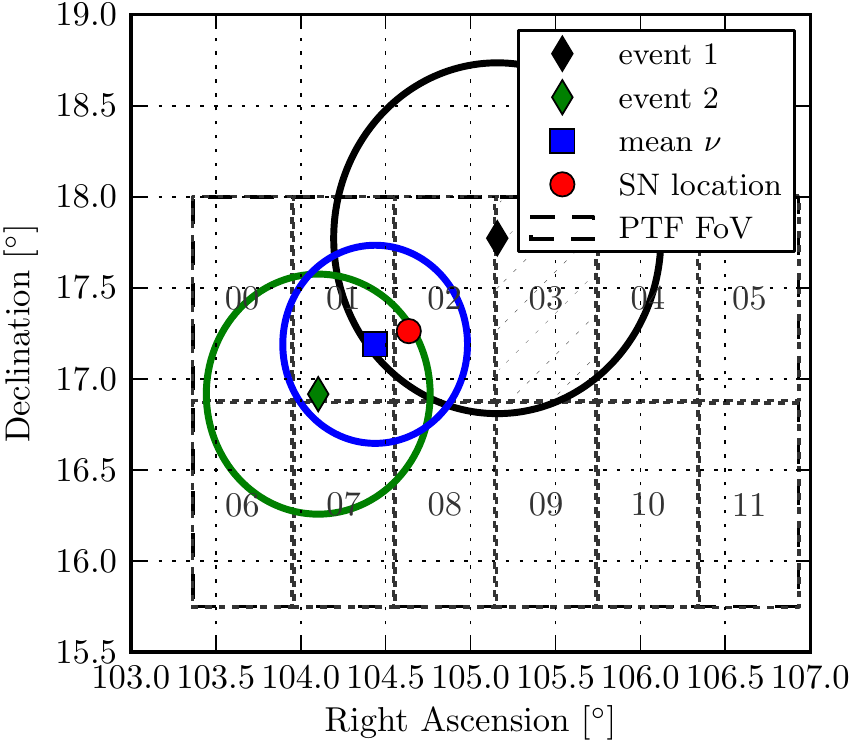}}
\caption{\emph{Left}: PTF reference-subtracted discovery image of the CCSN \emph{PTF12csy} taken on April 9, 2012 at the location of an IceCube neutrino doublet detected on March 30, 2012. \emph{Right:} Map showing the two neutrino event positions and uncertainties compared to the field of view of the PTF camera CCDs. (From \cite{Aartsen:2015trq}.)}
\label{fig:supernova}
\end{figure}

The number of neutrino ``doublets'' observed so far by these programs agrees with the rate expected from the atmospheric neutrino background. Although not a statistically significant correlation ($2\sigma$), the capability of the OFU program to detect optical transients has been demonstrated by the discovery~\cite{Aartsen:2015trq} of the CCSN \emph{PTF12csy} (Fig.~\ref{fig:supernova}) in PTF follow-up observations of a neutrino doublet position. 

Results from \emph{Swift} observations performed as part of the IceCube XFU program are presented in \cite{Evans:2015qia}. Seven 1-2 ks exposures are required to cover the $\sim0.5^{\circ}$ muon error circle with the $0.4^{\circ}$ XRT field-of-view. Given their increased sensitivity, XFU observations performed so far have unveiled more than 100 previously uncatalogued X-ray sources, although none of them appear to be clear counterparts for the neutrino events. These studies would greatly benefit from a deeper all-sky X-ray catalog, such as the one to be created by the \emph{eRosita}\cite{Merloni:2012uf} mission, which could be compared to new sources detected in triggered observations. The sensitivity to short transients would also be improved by the operation of an all-sky (or large field-of-view) X-ray telescope, which would reduce the hour-scale delay between the neutrino trigger and the start of X-ray observations.

Results from a similar X-ray and optical follow-up program in ANTARES are covered in~\cite{Adrian-Martinez:2015nin}. The Telescopes-ANTARES Target of Opportunity (TAToO) alert system~\cite{Ageron:2011pe} started operations in 2009, and it has triggered optical observations using the TAROT and ROTSE telescopes, and in X-rays with \emph{Swift} XRT. Also in this case, no significant optical or X-ray~\cite{Adrian-Martinez:2015nin} counterparts to the neutrino events have been detected so far. On September 1, 2015, \emph{Swift} observations triggered by an ANTARES alert revealed a variable, and previously unknown X-ray source in the 18-arcmin error circle of the neutrino event \footnote{Coincidentally, the ANTARES trigger position was found in the vicinity of the star Antares ($\alpha$ Sco).}. While this detection appears to have been caused by a chance alignment with an X-ray source, the strong multiwavelength follow-up campaign triggered by this detection highlights the interest of the astrophysical community in contributing to the discovery of the first neutrino source.

\section{Cosmic rays}\label{sec:cosmic-ray}

Cosmic rays are scattered by galactic (GMF) and intergalactic (IGMF) magnetic fields during propagation, limiting the applicability of spatial correlation studies with neutrino positions to the ultra-high energy cosmic ray range (UHECRs, $E_{\mathrm{CR}} \gtrsim 10^{18}$ eV). Simulations indicate that $10^{20}$ eV protons are deflected only a few degrees by the galactic magnetic field (GMF) during propagation, although significant uncertainties remain on this figure given our incomplete knowledge of the chemical composition of the UHECR flux and the strength and structure of the GMF and IGMF. If neutrons are present in the cosmic-ray flux, their limited decay range of $\sim 900$ kpc at $10^{20}$ eV restricts the reach of neutrino-neutron correlation searches to our immediate galactic vicinity. The range of UHECRs with energies above $10^{20}$ eV is also limited to a few tens of Mpc by the Greisen-Zatsepin-Kuzmin (GZK) energy-loss mechanism. 

The two main facilities dedicated to the study of UHECRs are the Pierre Auger Observatory (Auger)~\cite{Aab:2015zoa} in Mendoza, Argentina, and the Telescope Array\cite{2003PThPS.151..206F} (TA) in Utah, USA, with instrumented areas of 3000 and 800 km$^2$, respectively. No strong evidence for UHECR point sources has been found so far, although a recent analysis~\cite{Abbasi:2014lda} of Telescope Array data shows hints of a $20^{\circ}$ ``hotspot'' in the northern sky at energies above $5.7 \times 10^{19}$ eV with a significance of $3.4 \sigma$. Possible indications of a dipole anisotropy have been reported in Auger data at the $4\sigma$ level~\cite{ThePierreAuger:2014nja}.

A neutrino-UHECR connection is favored by the similarity between the astrophysical neutrino flux level measured by IceCube and the Waxman-Bahcall (WB) flux~\cite{PhysRevD.59.023002}, which represents an upper bound on the neutrino flux from UHECR sources, assuming they accelerate protons that convert most of their energy to pions. However, IceCube data currently favors a softer spectral index~\cite{Aartsen:2015knd} than the $\propto E^{-2}$ spectrum assumed by the WB model. Extending the energy range of neutrino observations may elucidate the role that UHECRs play in the neutrino spectrum. As neutrinos carry about 5\% of the parent cosmic-ray proton energy, the TeV-PeV neutrino sample used for these searches is too low in energy to be directly produced in UHECR interactions, which are expected to be observed at $E_{\nu} > 10^{16}$ eV. Rather, the TeV-PeV neutrinos, including those associated with the astrophysical flux detected by IceCube, are used as tracers of cosmic-ray acceleration that could be responsible for UHECR emission. 

\begin{figure}[thb]
\center
\raisebox{-0.5\height}{\includegraphics[width=0.75\textwidth]{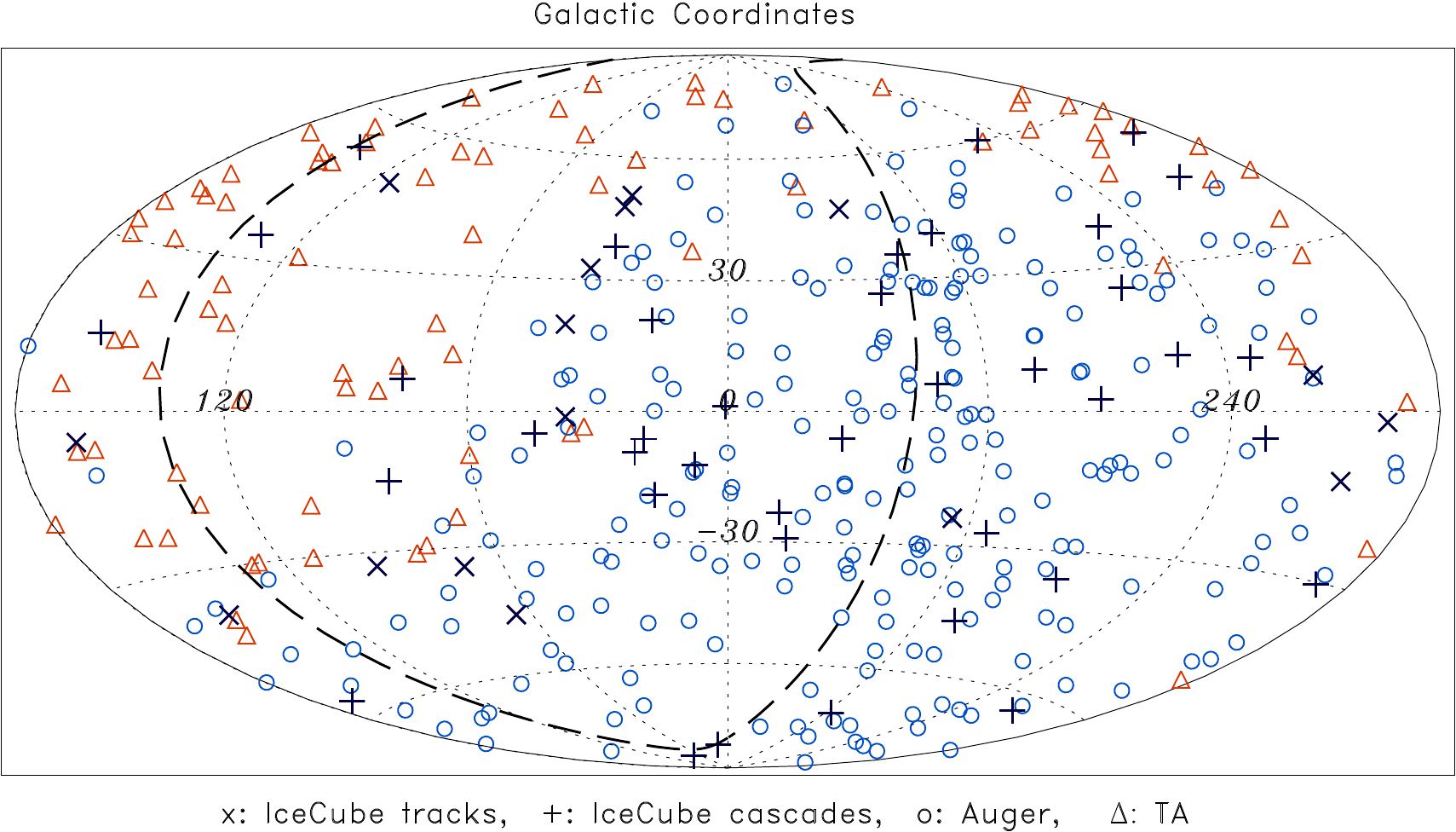}}
\caption{Galactic-coordinates sky map of IceCube high-energy cascades (plus signs), high-energy tracks (crosses), and UHECRs detected by Auger (circles) and TA (triangles). The Super-galactic plane is indicated by a dashed line. (From~\cite{Aartsen:2015dml}.)}
\label{fig:uhecr_map}
\end{figure}

Correlation studies using UHECR data from TA and Auger, and neutrinos events from ANTARES and IceCube, have been conducted in recent years, so far with no statistically significant detection. An ANTARES search~\cite{ANTARES:2012rcn} used a sample of up-going candidate neutrino events and Auger UHECR showers with energies above $5.5 \times 10^{19}$ eV. No significant correlation was found at several angular scales and an upper limit was derived on the neutrino flux from each UHECR direction. More recently, the IceCube, Auger, and TA collaborations presented results~\cite{Aartsen:2015dml} from an analysis that compared UHECR directions to both cascade and muon track neutrino events (Fig.~\ref{fig:uhecr_map}). No significant deviation from the isotropic expectation was found using high-energy muon tracks. For cascade events, a post-trial \emph{p}-value of $5 \times 10^{-4}$ was found for a typical angular separation of $22^{\circ}$ from UHECR events. An \emph{a posteriori} test where the UHECR positions were fixed gives a \emph{p}-value of $8.5 \times 10^{-3}$. The statistical significance appears to be driven by event pairs in the region of the TA ``hotspot'' and it will be interesting to follow how the correlation evolves as the data set of both UHECRs and high-energy neutrinos continues to grow over the coming years. 

Further enhancements to Auger~\cite{Aab:2015bza}, the planned expansion of TA to almost 2600 km$^{2}$, and the future launch of the JEM-EUSO mission~\cite{Ebisuzaki:2010zz} will provide a significant boost to UHECR statistics above $ 10^{20}$ eV, where GMF deflections are reduced. Joint searches using data from these upgraded instruments and next-generation neutrino telescopes will help explore current hints of an UHECR-neutrino correlation.

\section{Gravitational waves}\label{sec:gw}

The recent announcement of the first detection of gravitational waves by LIGO~\cite{PhysRevLett.116.061102} represents a groundbreaking result that opens yet another channel to study the extreme universe in addition to photons, cosmic rays, and neutrinos. It has been proposed that energetic sources such as GRBs, CCSNe, and soft gamma repeaters (SGR) are emitters of neutrinos and gravitational waves (GW). While the neutrinos are produced by particle interactions in the relativistic outflows from these objects, GWs are related to the bulk matter dynamics of the source progenitor. Searches for spatial and temporal correlations between GW and neutrino signals have a higher sensitivity to these type of sources than those performed separately on each channel. A combined study also enables searches for more exotic phenomena such as ``choked'' GRBs, where the jet is not able to break out from the progenitor and therefore no gamma rays are emitted. Other events that are too faint or obscured to be detected by EM telescopes, or that are missed by the limited sky coverage of these detectors, may also be observed through this approach.  Besides providing insights on the GW source progenitor, a coincident GW-neutrino detection can drastically shrink the source confidence region from several hundreds of square degrees (as in the case of the first GW detection) to the sub-square-degree level for neutrinos enabling targeted EM follow-up observations. The status of combined neutrino-GW searches leading to the first GW detection is presented in \cite{Ando:2012hna}.

The most sensitive GW observatories currently in operation are km-scale Michelson laser interferometers. The LIGO observatory~\cite{0034-4885-72-7-076901} operates detectors in two locations in the USA: Livingston, Louisiana, and Hanford, Washington. Both detectors were recently upgraded and in September 2015 the observatory started science operations for its ``Advanced LIGO'' (aLIGO) phase~\cite{0264-9381-27-8-084006}, during which the first GW signals were detected. A LIGO site in India has also been proposed, which would significantly improve the ability of the observatory to locate sources in the sky. In Europe, the Virgo detector (near Cascina, Italy) is currently being upgraded~\cite{advvirgo} to its ``Advanced Virgo'' configuration (AdV), while the GEO 600 observatory~\cite{Willke:2002bs} (near Sarstedt, Germany) is being used as a test-bench for advanced technology concepts. KAGRA~\cite{Somiya:2011np}, a future observatory in Kamioka, Japan, is currently under construction and the beginning of scientific operations is expected towards the end of the decade. 

\begin{figure}[htb]
\begin{center}
\raisebox{-0.5\height}{\includegraphics[width=0.75\textwidth]{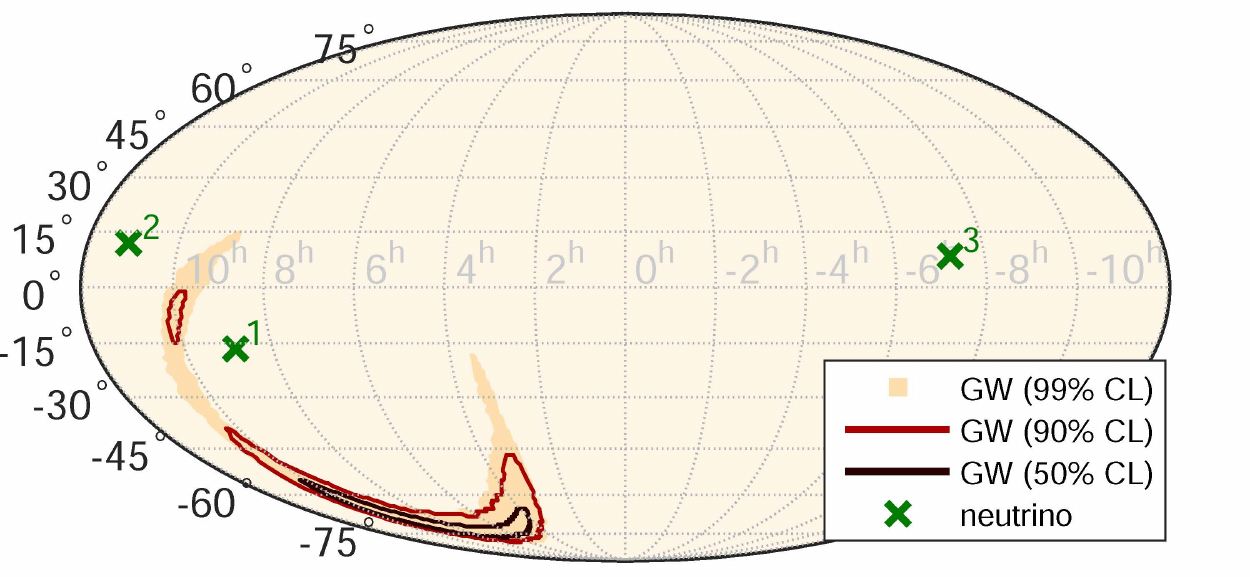}}
\end{center}
\caption{ Error regions for the GW event \emph{GW150914} for 50\%, 90\%, and 99\% containment levels with three IceCube neutrinos detected in a $\pm 500$s time window around the GW event~\cite{PhysRevLett.116.061102}.}
\label{fig:gw}
\end{figure}

Previous to the GW discovery, several searches for GW-neutrino spatial and temporal correlations were performed using GW data from  LIGO and Virgo, and neutrino events from ANTARES~\cite{AdrianMartinez:2012tf} and IceCube~\cite{Aartsen:2014mfp}, with no significant correlations detected. The first GW event, named \emph{GW150914}, was detected by LIGO on September 14th, 2015 and was produced by the merger of two $\sim30 \; M_{\astrosun}$ black holes at a distance of 410 Mpc. Although no neutrino signal is expected in a black hole-black hole merger, a search for coincident neutrino events was performed using data from ANTARES and IceCube. Three neutrino events (all from IceCube) were found in an \emph{a priori}-defined $\pm 500$ s window around the GW detection (Fig.~\ref{fig:gw}), in good agreement with atmospheric background expectations. Additionally, none of the neutrinos were spatially coincident with the GW uncertainty region and all-sky upper limits were derived on any potential associated neutrino source. 

A second LIGO GW event was recently announced~\cite{PhysRevLett.116.241103}, also detected during the first run of the aLIGO configuration, while a second run is expected to start towards the end of 2016. In addition to the continuing operation of aLIGO, the start of operations of AdV, the completion of KAGRA, the possible construction of the LIGO India site and the outstanding performance of the LISA Pathfinder mission~\cite{PhysRevLett.116.231101} promise a bright future for GW astronomy and for correlation studies with neutrinos. 

\section{Transient searches}\label{sec:transient}

Most transient searches described so far have been performed on the basis of individual agreements between neutrino observatories and one or more multi-messenger detectors. These studies have been aimed at exploring a particular detection channel or augmenting the sensitivity to certain types of sources. However, as the sources of astrophysical neutrinos remain unknown, a better approach is to combine all available measurements in order to search for temporal and spatial correlations. A realtime detection of an interesting correlation between two or more channels can be used to trigger follow-up observations using pointed instruments. The superior angular and energy resolution of these instruments can increase the significance of a potential transient signal. 

The Astrophysical Multimessenger Observatory Network (AMON)\footnote{http://amon.gravity.psu.edu/} is the current effort to realize this strategy. AMON provides the computational infrastructure to interconnect neutrino and gravitational-wave observatories with EM partners so that transient alerts can be transmitted without delay. A database of past alerts also enables archival coincidence searches of multi-messenger signals. 

Observatories are classified as ``triggering'' or ``follow-up'' instruments. Triggering instruments, such as IceCube and HAWC, have large fields of view and high duty cycles. These observatories send event information to AMON including the event detection time, its position and uncertainty, an estimate of the false-trigger probability associated with this event, and additional detector-dependent quantities. AMON will forward interesting triggers (such as PeV neutrino event locations) directly to follow-up observatories, while lower-significance ``sub-threshold'' events will be used by the AMON online analysis to search for realtime coincidences and issue alerts if a correlation is found. The distribution of the alerts is performed using the Gamma-ray Coordinates Network (GCN).

AMON has recently started realtime operations~\cite{AMON:ICRC}, and analyses of archival data sets made public by different partners have been performed~\cite{Keivani:2015kna}. For EM follow-ups of neutrinos associated with GRBs, a $>$1000 increase in efficiency is expected from conducting correlation studies between single neutrino events and several EM streams with respect to the current EM follow-up ``status quo'' that requires significant detections in each channel. 
\section{Conclusions and outlook}\label{sec:conclusions}

The high-energy astrophysical neutrino flux revealed by IceCube opens exciting possibilities to explore the extreme universe using multiple cosmic messengers. The impact of this breakthrough result has sparked a large number of observational programs
and theoretical studies aimed at unveiling the sources of astrophysical neutrinos using a multi-messenger approach which involves photons, cosmic rays
and gravitational waves. 

An optimal scenario for multi-messenger searches consists of several neutrino telescopes with multi-km$^3$ effective volumes that are capable of detecting a large number of astrophysical events, reconstructing them with good angular resolution, and broadcasting their positions in near-realtime to partner multi-messenger observatories. 
Work is currently underway in different fronts to achieve many of these desired goals. Two next-generation neutrino telescopes are in the works: IceCube-Gen2~\cite{Aartsen:2014njl}, a 10 km$^{3}$ extension to IceCube, and KM3NeT, the first km$^{3}$-scale detector to be built in the northern hemisphere. The joint operation of both detectors will greatly improve our sensitivity to neutrino point-sources, while providing a large high-purity astrophysical neutrino sample to use in follow-up observations. Even for current generation instruments, the sensitivity of counterpart searches can be significantly improved by exploiting the high-energy through-going muons detected by IceCube, which constitute the ``golden channel'' for multi-messenger searches given their sub-degree angular resolution and high astrophysical probability. Significant improvements in reconstruction techniques over the coming years will boost the angular resolution of cascade events in IceCube from its current value of $\sim15^{\circ}$, and it is feasible that KM3NeT will be able to deliver an angular resolution for these events at the level of $\sim2^{\circ}$. 

While the directional uncertainty for cascades prevents most follow-up observations from using pointed instruments given their limited sky coverage, large field of view detectors with high-duty cycles (such as \emph{Fermi}-LAT and HAWC in gamma rays, Auger and TA in cosmic rays, and the aLIGO and AdV in gravitational waves) can  be used to search for temporal and spatial correlations. The sensitivity to counterparts will continue to increase thanks to the construction and operation of next-generation instruments, such as CTA, LHAASO, and HiScore (gamma-rays); JEM-EUSO (cosmic rays); and KAGRA (gravitational waves). The last remaining step is to interconnect this vast observational network. In this sense, the AMON network is currently starting operations and will provide an avenue for the rapid dissemination of neutrino alerts.

In summary, we foresee that over the next few years a greatly-improved global network of multi-messenger observatories will enter regular operations, and we look forward to the revolutionary discovery of the first point sources of astrophysical neutrinos as the crowning achievement for this joint international effort.
\section{Acknowledgements}\label{sec:acknow}

The author would like to thank Imre Bartos, Jon Dumm, Geraldina Golup, Jamie Holder, Azadeh Keivani, Thomas Kintscher, Gernot Maier, Aurore Mathieu, Reshmi Mukherjee, Daniel Nieto, Michael Shaevitz, Fabian Sch\"{u}ssler, Ignacio Taboada, Gordana Te{\v s}i{\'c} and Stefan Westerhoff for helpful discussions and comments, and for providing some of the material used in this work. 

The author acknowledges support from the US National Science Foundation through NSF grant PHYS-1505811.

\bibliographystyle{abbrv}
\bibliography{multimessenger}

\begin{thebibliography}{10}

\bibitem{Kelner:2006tc}
S.~R. Kelner, Felex~A. Aharonian, and V.~V. Bugayov.
\newblock {Energy spectra of gamma-rays, electrons and neutrinos produced at
  proton-proton interactions in the very high energy regime}.
\newblock {\em Phys. Rev.}, D74:034018, 2006.
\newblock [Erratum: Phys. Rev.D79,039901(2009)].

\bibitem{Kelner:2008ke}
S.~R. Kelner and F.~A. Aharonian.
\newblock {Energy spectra of gamma-rays, electrons and neutrinos produced at
  interactions of relativistic protons with low energy radiation}.
\newblock {\em Phys. Rev.}, D78:034013, 2008.
\newblock [Erratum: Phys. Rev.D82,099901(2010)].

\bibitem{HESE3yr}
M.~G. Aartsen et~al.
\newblock {Observation of High-Energy Astrophysical Neutrinos in Three Years of
  IceCube Data}.
\newblock {\em Phys. Rev. Lett.}, 113:101101, Sep 2014.

\bibitem{IC4yearPS}
M.~G. Aartsen et~al.
\newblock {Searches for Extended and Point-like Neutrino Sources with Four
  Years of IceCube Data}.
\newblock {\em ArXiv e-prints}, June 2014.

\bibitem{2015ApJ...805L...5A}
M.~G. {Aartsen} et~al.
\newblock {Search for Prompt Neutrino Emission from Gamma-Ray Bursts with
  IceCube}.
\newblock {\em ApJL}, 805:L5, May 2015.

\bibitem{I3MoonShadow}
M.~G. Aartsen et~al.
\newblock Observation of the cosmic-ray shadow of the moon with icecube.
\newblock {\em Phys. Rev. D}, 89:102004, May 2014.

\bibitem{AdrianMartinez:2012rp}
S.~Adrian-Martinez et~al.
\newblock {Search for Cosmic Neutrino Point Sources with Four Year Data of the
  ANTARES Telescope}.
\newblock {\em Astrophys. J.}, 760:53, 2012.

\bibitem{Margiotta:2014gza}
Annarita Margiotta.
\newblock {The KM3NeT deep-sea neutrino telescope}.
\newblock {\em Nucl. Instrum. Meth.}, A766:83--87, 2014.

\bibitem{KM3NetCascades}
{{D.~Stransky et al. (for the KM3NeT Collaboration)}}.
\newblock {Reconstruction of cascade-type neutrino events in KM3NeT/ARCA}.
\newblock {\em PoS}, ICRC2015:1108, 2015.

\bibitem{Ahlers:2013xia}
Markus Ahlers and Kohta Murase.
\newblock {Probing the Galactic Origin of the IceCube Excess with Gamma-Rays}.
\newblock {\em Phys. Rev.}, D90(2):023010, 2014.

\bibitem{Kistler:2015oae}
Matthew~D. Kistler.
\newblock {On TeV Gamma Rays and the Search for Galactic Neutrinos}.
\newblock 2015.

\bibitem{Taylor:2014hya}
Andrew~M. Taylor, Stefano Gabici, and Felix Aharonian.
\newblock {Galactic halo origin of the neutrinos detected by IceCube}.
\newblock {\em Phys. Rev.}, D89(10):103003, 2014.

\bibitem{Lunardini:2015laa}
Cecilia Lunardini, Soebur Razzaque, and Lili Yang.
\newblock {Multimessenger study of the Fermi bubbles: Very high energy gamma
  rays and neutrinos}.
\newblock {\em Phys. Rev.}, D92(2):021301, 2015.

\bibitem{IceCube:2012nn}
R.~Abbasi et~al.
\newblock {IceTop: The surface component of IceCube}.
\newblock {\em Nucl. Instrum. Meth.}, A700:188--220, 2013.

\bibitem{Abeysekara:2013tza}
A.~U. Abeysekara et~al.
\newblock {Sensitivity of the High Altitude Water Cherenkov Detector to Sources
  of Multi-TeV Gamma Rays}.
\newblock {\em Astropart. Phys.}, 50-52:26--32, 2013.

\bibitem{Cui:2014bda}
Shuwang Cui, Ye~Liu, Yujuan Liu, and Xinhua Ma.
\newblock {Simulation on gamma ray astronomy research with LHAASO-KM2A}.
\newblock {\em Astropart. Phys.}, 54:86--92, 2014.

\bibitem{Tluczykont:2014cva}
M.~Tluczykont et~al.
\newblock {The HiSCORE concept for gamma-ray and cosmic-ray astrophysics beyond
  10 TeV}.
\newblock {\em Astropart. Phys.}, 56:42--53, 2014.

\bibitem{2009ApJ...697.1071A}
W.~B. {Atwood} et~al.
\newblock {The Large Area Telescope on the Fermi Gamma-Ray Space Telescope
  Mission}.
\newblock {\em ApJ}, 697:1071--1102, June 2009.

\bibitem{Aharonian:2006pe}
F.~Aharonian et~al.
\newblock {Observations of the Crab Nebula with H.E.S.S}.
\newblock {\em Astron. Astrophys.}, 457:899--915, 2006.

\bibitem{2013arXiv1309.6979I}
M.~G. {Aartsen} et~al.
\newblock {The IceCube Neutrino Observatory Part I: Point Source Searches}.
\newblock {\em ArXiv e-prints}, September 2013.

\bibitem{2002APh....17..221W}
T.~C. {Weekes} et~al.
\newblock {VERITAS: the Very Energetic Radiation Imaging Telescope Array
  System}.
\newblock {\em Astroparticle Physics}, 17:221--243, May 2002.

\bibitem{FermiLATPerformance}
{{Fermi-LAT Collaboration}}.
\newblock {Fermi-LAT Performance}.
\newblock
  \url{http://www.slac.stanford.edu/exp/glast/groups/canda/lat_Performance.htm},
  2015.

\bibitem{Atwood:2013rka}
W.~Atwood et~al.
\newblock {Pass 8: Toward the Full Realization of the Fermi-LAT Scientific
  Potential}.
\newblock {\em 2012 Fermi Symposium proceedings - eConf C121028}, 2013.

\bibitem{Park:2015ysa}
Nahee Park.
\newblock {Performance of the VERITAS experiment}.
\newblock In {\em {Proceedings, 34th International Cosmic Ray Conference (ICRC
  2015)}}, 2015.

\bibitem{2016APh....72...76A}
J.~{Aleksi{\'c}} et~al.
\newblock {The major upgrade of the MAGIC telescopes, Part II: A performance
  study using observations of the Crab Nebula}.
\newblock {\em Astroparticle Physics}, 72:76--94, January 2016.

\bibitem{Holler:2015tca}
M.~Holler, A.~Balzer, R.~Chalmé-Calvet, M.~de~Naurois, and D.~Zaborov.
\newblock {Photon Reconstruction for H.E.S.S. Using a Semi-Analytical Shower
  Model}.
\newblock In {\em {Proceedings, 34th International Cosmic Ray Conference (ICRC
  2015)}}, 2015.

\bibitem{CTASensitivity}
{{CTA Consortium}}.
\newblock {CTA Performance}.
\newblock \url{https://portal.cta-observatory.org/Pages/CTA-Performance.aspx},
  2015.

\bibitem{LHAASO:ICRC}
{He, H. and others}.
\newblock {Design highlights and status of the LHAASO project}.
\newblock {\em PoS}, ICRC2015:1010, 2015.

\bibitem{Ackermann:2014usa}
M.~Ackermann et~al.
\newblock {The spectrum of isotropic diffuse gamma-ray emission between 100 MeV
  and 820 GeV}.
\newblock {\em Astrophys. J.}, 799:86, 2015.

\bibitem{DiMauro:2015tfa}
Mattia Di~Mauro and Fiorenza Donato.
\newblock {Composition of the Fermi-LAT isotropic gamma-ray background
  intensity: Emission from extragalactic point sources and dark matter
  annihilations}.
\newblock {\em Phys. Rev.}, D91(12):123001, 2015.

\bibitem{Ando:2015bva}
Shin'ichiro Ando, Irene Tamborra, and Fabio Zandanel.
\newblock {Tomographic Constraints on High-Energy Neutrinos of Hadronuclear
  Origin}.
\newblock {\em Phys. Rev. Lett.}, 115(22):221101, 2015.

\bibitem{Bechtol:2015uqb}
Keith Bechtol, Markus Ahlers, Mattia Di~Mauro, Marco Ajello, and Justin
  Vandenbroucke.
\newblock {Evidence against star-forming galaxies as the dominant source of
  IceCube neutrinos}.
\newblock 2015.

\bibitem{Brown21072015}
Anthony~M. Brown, Jenni Adams, and Paula~M. Chadwick.
\newblock $\gamma$-ray observations of extraterrestrial neutrino track events.
\newblock {\em Monthly Notices of the Royal Astronomical Society},
  451(1):323--331, 2015.

\bibitem{LATBigBird}
M.~Kadler et~al.
\newblock Coincidence of a high-fluence blazar outburst with a pev-energy
  neutrino event.
\newblock {\em Nat Phys}, advance online publication:--, 04 2016.

\bibitem{Schussler:2013lja}
F.~Schüssler et~al.
\newblock {Multiwavelength study of the region around the ANTARES neutrino
  excess}.
\newblock In {\em {Proceedings, 33rd International Cosmic Ray Conference
  (ICRC2013)}}, 2013.

\bibitem{2015arXiv150900517S}
{M. Santander, for the VERITAS and IceCube Collaborations}.
\newblock {Searching for TeV gamma-ray emission associated with IceCube
  high-energy neutrinos using VERITAS}.
\newblock {\em ArXiv e-prints}, September 2015.

\bibitem{2015arXiv150903035S}
{{F. Sch{\"u}ssler} et al. for the H.~E.~S.~S.~Collaboration}.
\newblock {The H.E.S.S. multi-messenger program}.
\newblock {\em ArXiv e-prints}, September 2015.

\bibitem{Aartsen:2015rwa}
M.~G. Aartsen et~al.
\newblock {Evidence for Astrophysical Muon Neutrinos from the Northern Sky with
  IceCube}.
\newblock {\em Phys. Rev. Lett.}, 115(8):081102, 2015.

\bibitem{PeVmuonATel}
{{L. R\"adel and S. Schoenen for the IceCube Collaboration}}.
\newblock {ATel \#7856}.
\newblock \url{http://www.astronomerstelegram.org/?read=7856}, 2015.

\bibitem{Acharya:2013sxa}
B.~S. Acharya et~al.
\newblock {Introducing the CTA concept}.
\newblock {\em Astropart. Phys.}, 43:3--18, 2013.

\bibitem{Moiseev:2015lva}
A.~A. Moiseev et~al.
\newblock {Compton-Pair Production Space Telescope (ComPair) for MeV Gamma-ray
  Astronomy}.
\newblock 2015.

\bibitem{Abbasi:2011ja}
R.~Abbasi et~al.
\newblock {Searching for soft relativistic jets in Core-collapse Supernovae
  with the IceCube Optical Follow-up Program}.
\newblock {\em Astron. Astrophys.}, 539:A60, 2012.

\bibitem{Kowalski:2007xb}
Marek Kowalski and Anna Mohr.
\newblock {Detecting neutrino-transients with optical follow-up observations}.
\newblock {\em Astropart. Phys.}, 27:533--538, 2007.

\bibitem{Aartsen:2015trq}
M.~G. Aartsen et~al.
\newblock {Detection of a Type IIn Supernova in Optical Follow-up Observations
  of IceCube Neutrino Events}.
\newblock {\em Astrophys. J.}, 811(1):52, 2015.

\bibitem{Evans:2015qia}
P.~A. Evans et~al.
\newblock {Swift follow-up of IceCube triggers, and implications for the
  Advanced-LIGO era}.
\newblock {\em Mon. Not. Roy. Astron. Soc.}, 448(3):2210--2223, 2015.

\bibitem{Merloni:2012uf}
A.~Merloni et~al.
\newblock {eROSITA Science Book: Mapping the Structure of the Energetic
  Universe}.
\newblock {\em MPE document (online)}, 2012.

\bibitem{Adrian-Martinez:2015nin}
S.~Adrian-Martinez et~al.
\newblock {Optical and X-ray early follow-up of ANTARES neutrino alerts}.
\newblock {\em arXiv preprint}, 2015.

\bibitem{Ageron:2011pe}
M.~Ageron et~al.
\newblock {The ANTARES Telescope Neutrino Alert System}.
\newblock {\em Astropart. Phys.}, 35:530--536, 2012.

\bibitem{Aab:2015zoa}
Alexander Aab et~al.
\newblock {The Pierre Auger Cosmic Ray Observatory}.
\newblock {\em Nucl. Instrum. Meth.}, A798:172--213, 2015.

\bibitem{2003PThPS.151..206F}
M.~{Fukushima}.
\newblock {Telescope Array Project for Extremely High Energy Cosmic Rays}.
\newblock {\em Progress of Theoretical Physics Supplement}, 151:206--210, 2003.

\bibitem{Abbasi:2014lda}
R.~U. Abbasi et~al.
\newblock {Indications of Intermediate-Scale Anisotropy of Cosmic Rays with
  Energy Greater Than 57 EeV in the Northern Sky Measured with the Surface
  Detector of the Telescope Array Experiment}.
\newblock {\em Astrophys. J.}, 790:L21, 2014.

\bibitem{ThePierreAuger:2014nja}
Alexander Aab et~al.
\newblock {Large Scale Distribution of Ultra High Energy Cosmic Rays Detected
  at the Pierre Auger Observatory With Zenith Angles up to 80°}.
\newblock {\em Astrophys. J.}, 802(2):111, 2015.

\bibitem{PhysRevD.59.023002}
Eli Waxman and John Bahcall.
\newblock High energy neutrinos from astrophysical sources: An upper bound.
\newblock {\em Phys. Rev. D}, 59:023002, Dec 1998.

\bibitem{Aartsen:2015knd}
M.~G. Aartsen et~al.
\newblock {A combined maximum-likelihood analysis of the high-energy
  astrophysical neutrino flux measured with IceCube}.
\newblock {\em Astrophys. J.}, 809(1):98, 2015.

\bibitem{ANTARES:2012rcn}
S.~Adri\'{a}n-Martínez et~al.
\newblock {Search for a correlation between ANTARES neutrinos and Pierre Auger
  Observatory UHECRs arrival directions}.
\newblock {\em Astrophys. J.}, 774:19, 2013.

\bibitem{Aartsen:2015dml}
M.~G. Aartsen et~al.
\newblock {Search for correlations between the arrival directions of IceCube
  neutrino events and ultrahigh-energy cosmic rays detected by the Pierre Auger
  Observatory and the Telescope Array}.
\newblock {\em JCAP}, 1601(01):037, 2016.

\bibitem{Aab:2015bza}
Alexander Aab et~al.
\newblock {The Pierre Auger Observatory: Contributions to the 34th
  International Cosmic Ray Conference (ICRC 2015)}.
\newblock {\em PoS}, ICRC2015:593, 2015.

\bibitem{Ebisuzaki:2010zz}
Toshikazu Ebisuzaki et~al.
\newblock {The JEM-EUSO mission to explore the extreme universe}.
\newblock {\em AIP Conf. Proc.}, 1238:369--376, 2010.

\bibitem{PhysRevLett.116.061102}
B.~P. Abbott et~al.
\newblock Observation of gravitational waves from a binary black hole merger.
\newblock {\em Phys. Rev. Lett.}, 116:061102, Feb 2016.

\bibitem{Ando:2012hna}
Shin'ichiro Ando et~al.
\newblock {Colloquium: Multimessenger astronomy with gravitational waves and
  high-energy neutrinos}.
\newblock {\em Rev. Mod. Phys.}, 85(4):1401--1420, 2013.

\bibitem{0034-4885-72-7-076901}
B~P Abbott et~al.
\newblock Ligo: the laser interferometer gravitational-wave observatory.
\newblock {\em Reports on Progress in Physics}, 72(7):076901, 2009.

\bibitem{0264-9381-27-8-084006}
Gregory~M Harry and the LIGO Scientific~Collaboration.
\newblock {Advanced LIGO: the next generation of gravitational wave detectors}.
\newblock {\em Classical and Quantum Gravity}, 27(8):084006, 2010.

\bibitem{advvirgo}
{Virgo Collaboration}.
\newblock {Advanced Virgo Baseline Design}.
\newblock Report No. VIR-027A-09, 2009.

\bibitem{Willke:2002bs}
B.~Willke et~al.
\newblock {The GEO 600 gravitational wave detector}.
\newblock {\em Class. Quant. Grav.}, 19:1377--1387, 2002.

\bibitem{Somiya:2011np}
Kentaro Somiya.
\newblock {Detector configuration of KAGRA: The Japanese cryogenic
  gravitational-wave detector}.
\newblock {\em Class. Quant. Grav.}, 29:124007, 2012.

\bibitem{AdrianMartinez:2012tf}
S.~Adrian-Martinez et~al.
\newblock {A First Search for coincident Gravitational Waves and High Energy
  Neutrinos using LIGO, Virgo and ANTARES data from 2007}.
\newblock {\em JCAP}, 1306:008, 2013.

\bibitem{Aartsen:2014mfp}
M.~G. Aartsen et~al.
\newblock {Multimessenger search for sources of gravitational waves and
  high-energy neutrinos: Initial results for LIGO-Virgo and IceCube}.
\newblock {\em Phys. Rev.}, D90(10):102002, 2014.

\bibitem{PhysRevLett.116.241103}
B.~P. Abbott et~al.
\newblock Gw151226: Observation of gravitational waves from a 22-solar-mass
  binary black hole coalescence.
\newblock {\em Phys. Rev. Lett.}, 116:241103, Jun 2016.

\bibitem{PhysRevLett.116.231101}
M.~Armano et~al.
\newblock Sub-femto-$g$ free fall for space-based gravitational wave
  observatories: Lisa pathfinder results.
\newblock {\em Phys. Rev. Lett.}, 116:231101, Jun 2016.

\bibitem{AMON:ICRC}
G.~{Te{\v s}i{\'c}} and A.~Keivani.
\newblock {AMON: Transition to Real-Time Operations}.
\newblock {\em PoS}, ICRC2015:329, 2015.

\bibitem{Keivani:2015kna}
A.~{Keivani}, D.~B. {Fox}, G.~{Te{\v s}i{\'c}}, D.~F. {Cowen}, and
  J.~{Fixelle}.
\newblock {AMON Searches for Jointly-Emitting Neutrino + Gamma-Ray Transients}.
\newblock {\em PoS}, ICRC2015:786, 2015.

\bibitem{Aartsen:2014njl}
M.~G. Aartsen et~al.
\newblock {IceCube-Gen2: A Vision for the Future of Neutrino Astronomy in
  Antarctica}.
\newblock {\em arXiv preprint}, 2014.

\end{thebibliography}

\end{document}